# Neutron-diffraction study of field-induced transitions in the heavy-fermion compound $Ce_2RhIn_8$

E. G. Moshopoulou[a*], K. Prokes[b], E. Garcia-Matres[b], P. G. Pagliuso[c], J. L. Sarrao[c], and J. D. Thompson[c]

[a]*National Center for Scientific Research "Demokritos", Institute of Materials Science, 15310 Agia Paraskevi, Greece*
[b]*Hahn-Meitner-Institute, SF-2, Glienicker Strasse 100, D-14109 Berlin, Germany*
[c]*Condensed Matter and Thermal Physics, Los Alamos National Laboratory, MS K764, Los Alamos NM 87545, U.S.A.*

(February 10, 2002)

**Abstract**

We present neutron diffraction measurements in high magnetic fields (0 to 14.5 T) and at low temperatures (2.5, 2.3, 0.77 and 0.068 K) on single crystals of the tetragonal heavy fermion antiferromagnet $Ce_2RhIn_8$. For $B//[110]$ the field dependence of selected magnetic and nuclear reflections reveals that the material undergoes several transitions, the temperature dependence of which suggests a complex B-T phase diagram. We present the detailed evolution of the integrated intensities of selected reflections and discuss the associated field-induced transitions.

# 1. Introduction

Heavy Fermion (HF) materials exhibit fascinating low-temperature properties that are dramatically different from their ordinary-metal counterparts. The origin of their unusual properties is the strong interactions between the conduction electrons and the magnetic moments in these $f$-electron compounds and alloys. How to theoretically treat these strong many-body correlations is yet an unsolved problem and, despite considerable progress, more theoretical and experimental work is needed to unravel the behavior of these systems. One experimental strategy to probe the physics of HF materials and thus to suggest new directions to the theory is to consider the energy scales that govern their properties and use external "perturbation" such as a magnetic field whose characteristic energies are comparable with these scales. Magnetic field can continuously turn off some of the strong correlation effects in HF systems, influence the competition between their ground states and induce important metamagnetic transitions. The magnetic field needed is of the order of $k_B T^*/\mu_B$, where $T^*$ is the coherence temperature below which moment compensation occurs. Typically, the required field to probe the physics of HF is of the order of tens Tesla. Because combined high-fields/neutron-scattering facilities are scarce, very few studies have been performed to date with neutrons on field-induced transitions and consequently, for most HF systems, the origin and the microscopic details of such transitions remain unknown.

Recently, field-induced magnetic transitions have been observed in the HF antiferromagnet $Ce_2RhIn_8$ by field-dependent heat capacity experiments [1]. $Ce_2RhIn_8$ orders antiferromagnetically below Néel temperature $T_N$ = 2.8 K and has an enhanced value of electronic specific heat coefficient $\gamma$ ~ 400 mJ/mol-Ce $K^2$ (above $T_N$), which qualifies it as a new HF system [2]. Conventional powder x-ray diffraction showed that the material adopts the tetragonal $Ho_2CoGa_8$ type-structure [3] with space group P4/*mmm* and cell parameters a = 4.665(1) Å and c = 12.244(5) Å. Compounds of this type-structure are intergrowth structures built by bilayers of face-sharing distorted cuboctahedra ([CeIn$_3$] in the case of Ce$_2$RhIn$_8$) and monolayers of edge-sharing rectangular parallelepipeds ([RhIn$_2$] for Ce$_2$RhIn$_8$) alternatively stacked along the [001] direction. Given this structural arrangement, it is very likely that the magnetic structure of Ce$_2$RhIn$_8$ will be similar to that of the cubic parent compound CeIn$_3$, which is a collinear HF antiferromagnet. Indeed, Ce$_2$RhIn$_8$ adopts also a collinear antiferromagnetic structure with propagation vector (1/2,1/2,1) and a staggered moment of 0.55(6) µ$_B$ per Ce at 1.6 K, tilted 38(2)° from the *c*-axis [4]. The field-dependent specific heat experiments, mentioned above, were carried out using fields up to 10 T applied along the *c*- and *a*-axis of the crystal. For *B*//*c*, no field-induced transitions were found and $T_N$ and $\gamma$ decrease when the field increases in accord with several other HF materials [5]. However, when *B*//*a*, several field-induced transitions of both first and second order were observed. Recent magnetotransport, thermal expansion and magnetostriction measurements suggest also that Ce$_2$RhIn$_8$ exhibits a complicated *B-T* magnetic phase diagram [6].

In our experiments we used single crystal neutron diffraction in high magnetic fields (from 0 to 14.5 T) and at low temperatures (from 0.068 K to 2.5 K) to further investigate the interesting field-induced transitions inferred from the bulk measurements.

**2. Experimental**

The single-crystal neutron diffraction experiments were carried out using the two-axis diffractometer E4 at the Berlin Neutron Scattering Center in Hahn-Meitner Institute. The plane (002) of a pyrolithic graphite crystal was used as a focusing monochromator. The wavelength of the selected neutrons was 2.4432

Å. The neutron flux at the sample position was 1.8×10$^6$ n/cm$^2$s. The single crystal used was grown by the flux technique out of In flux. In order to be able to reach the strong (1/2,1/2,1) magnetic reflection of Ce$_2$RhIn$_8$ [4], we aligned the crystal with its [110] direction vertically. External magnetic field up to 14.5 T was provided by the vertical cryomagnet VM-1. This geometry, i.e. with *B*//[110], allows investigation of the field and temperature dependence of the magnetic reflection (1/2,1/2,1) as well as of the nuclear peaks (0,0,1) and (1,1,1). The study of nuclear reflections aims to find whether any broadening or splitting of these peaks occurs and if the field contributes intensity of magnetic origin to the nuclear intensity of these peaks.

Field-scans were taken at fixed temperatures below the ordering temperature $T_N$ = 2.8 K. In each case, the sample was heated well above $T_N$ and cooled in zero-field to the required temperature; then the field was increased in steps up to 14.5 T and finally it was decreased in bigger steps back to zero. We carried out such experiments at 2.5, 2.3, 0.77, 0.068 K for the reflections (1/2,1/2,1), (0,0,1) and (1,1,1) using the *ω*-scan mode. Measured peaks were fitted using the software "Bean" (BENSC Analysis) in order to deduce accurately its integrated intensity, peak position and full width at half-maximum (FWHM). Because at this stage of our study, we aimed to determine the field and temperature dependence of each reflection and not to compare different peaks or deduce the magnetic structures, it was not necessary to correct the measured intensities for the effects of absorption, extinction and *λ*/2 contamination.

3. Results

As a first step, the temperature dependent magnetic reflections (1/2,1/2,1), (1/2,1/2,2), (3/2,3/2,1) as well as several nuclear peaks were measured at zero-field. The relative intensities of the magnetic reflections were in agreement with the ones reported in reference 4. No intensity was observed at the reciprocal space point (1/2,1/2,0).

*3.1. Field-dependence of (1/2,1/2,1) at 2.5, 2.3, 0.77 and 0.068 K*

Fig. 1 shows the variation of the integrated intensity of the magnetic reflection (1/2,1/2,1) as a function of field at four fixed temperatures.

At 2.5 K (Fig. 1a), the intensity decreases rapidly as the field increases from 0 to 1 T and it is completely suppressed for $B \geq 2$ T up to 14.5 T. This rapid decrease of the intensity suggests that a rather first-order like transition should occur at the magnetic structure of $Ce_2RhIn_8$ between 1 and 2 T. Whether this transition is to a non-magnetic state because the field depresses $T_N$, remains to be verified. It is possible that only the propagation vector changes and above 2 T there is a completely new magnetic state. When the field decreases from 14.5 back to 0 T the variation of the integrated intensity shows perfect reversibility, thus proving that the observed behavior is an intrinsic property of the sample.

The evolution of the intensity as a function of field is dramatically different at the slightly lower temperature of 2.3 K (Fig. 1b). While more data are needed to determine accurately this evolution, these preliminary measurements reveal that the intensity increases slightly with increasing field reaching a maximum possibly at ≈ 12 T and then it decreases rather rapidly as the field rises to 14.5 T. These observations suggest a phase transition at high fields.

We now turn to much lower temperatures. At 0.77 K (Fig. 1c), field increase from 0 to ≈12 T induces a decrease of the magnetic intensity; then from ≈12 T to 14.5 T the magnetic intensity remains constant. Subsequent field decrease from 14.5 to 0 T induces an almost reversible variation of the magnetic intensity.

At 0.068 K (Fig. 1d), the evolution of the magnetic intensity as a function of the increasing field is quite similar as at 0.77 K but more pronounced. However, contrary to what observed at 0.77 K, as the field decreases from 14.5 back to 0 T, there is clear irreversibility at fields lower to ≈ 4 T. Possibly, at such low temperature the relaxation time of the system is longer than the time of measurement. Another possible explanation is that at 0.068 K, the increasing magnetic field causes growth of magnetic domains in the crystal and then as the field decreases irreversible domain repopulation takes place.

*3.2. Field-dependence of (1,1,1) at 2.5, 2.3, 0.77 and 0.068 K*

The field-dependence of the integrated intensity of the nuclear (1,1,1) reflection at 2.5 K is shown in Fig. 2. Field of 1 T induces a small magnetic component at the top of the zero-field nuclear intensity. This observation can be correlated with the rapid decrease of the magnetic peak (1/2,1/2,1) at 1 T, reinforcing the conclusion drawn previously that a magnetic transition occurs at 2.5 K and about 1 T. For fields higher than 2 T, the intensity remains almost equal to its initial zero-field value. At the other temperatures no variations of the intensity of (1,1,1) were detected.

*3.3. Field-dependence of (0,0,1) at 2.5, 2.3, 0.77 and 0.068 K*

Interestingly, at 2.5 K a field of 14.5 induces a slight increase of the integrated intensity of the nuclear peak (0,0,1). As shown in Fig. 3a, the intensity remains practically constant up to ≈ 11 T and then it rises as the field increases to 14.5 T. This magnetic contribution at the top of the nuclear intensity implies that at fields higher than ≈ 11 T, the material adopts a new magnetic state. Field decrease from 14.5 to 0 T induces a reversible decrease of the intensity, which, at zero-field, becomes almost equal to its initial zero-field value.

Contrary to the above behavior, at 2.3 K, the integrated intensity of the (0,0,1) reflection remains practically constant up to 14.5 T. At 0.77 K, (0,0,1) is also field-independent up to 14.5 T. At 0.068 K though, high fields induce a slight increase of the intensity reminiscent of the one observed at 2.5 K (Fig. 3b).

## 4. Discussion

The first results of neutron diffraction experiments on field-induced transitions in $Ce_2RhIn_8$ provide evidence of the very complicated microscopic picture of the *B-T* phase diagram of this material. Although more data are needed to determine the *B-T* phase diagram, the present experiments with *B*//[110] revealed several field induced transitions: At 2.5 K there is a transition between 1 and 2 T, which influences both the magnetic (1/2,1/2,1) and nuclear (1,1,1) reflections. This observation is in relative agreement with the specific heat measurements, which however were carried out with *B*//*a*, but also reveal a field-induced transition at 2.5 T. At this same temperature, one more transition occurs at higher fields (≈ 14.5 T) as evidenced by the change of the intensity of the nuclear peak (0,0,1) at high fields. Consequently at 2.5 K there are two different magnetic phases at low (< 2 T) and high (>13.5 T) fields and at least one phase for intermediate fields. Additional measurements in the region between 2 and 13 T are needed to find what transitions occur within this region. At 2.3 K, the field dependence of the magnetic structure is very different from the one at 2.5 K. While more data are needed, from the present measurements there is evidence of a phase transition at 2.3 K and about 12 T. Finally, at the lower temperatures of 0.77 and 0.068 K the low- and high-field magnetic structures are different and yet the high-field magnetic structure at 0.77 K differs from the one at 0.068 K.

As mentioned previously (Sec. 2), besides the integrated intensity two other parameters were deduced from the peak-fittings: the position and the FWHM. The fields used do not change visibly the position of the peaks (1/2,1/2,1), (0,0,1) and (1,1,1). However, they affect the FWHM's. At 2.5, 2.3 and 0.75 K, increase of the magnetic field causes a decrease of the FWHM of the magnetic (1/2,1/2,1) reflection and consequently an increase of the magnetic correlation length. At 0.068 K, the FWHM increases slightly with field up to 12 T and then it drops down for $B$=14.5 T, implying again that 14.5 T induce increase of the magnetic correlation length. As regards the nuclear reflection (0,0,1), which varies significantly with fields only at 2.5 K and 0.68 K, its FWHM increases at high fields for both these temperatures. This broadening can be due, at least in part, to strain introduced to the lattice by the high magnetic field. Slight broadening of the nuclear peak (1,1,1) was also observed at 2.5 K and 1 T. Besides these changes of the FWHM, no peak splitting could be observed.

These results of neutron diffraction experiments in magnetic fields add information to our present knowledge of field-induced transitions in $Ce_2RhIn_8$. Given the complexity of the $B$-$T$ phase diagram of this system, additional neutron diffraction experiments are clearly needed. It is particularly important to study the field dependence of more reflections especially in the temperature range 0.8 and 2.4 K and carry out scans along various reciprocal directions to search for field-induced reflections. From such data the magnetic structure of the various magnetic phases will be deduced. Higher resolution data would also be helpful to establish the absence of magnetically driven structural distortions suggested by the present data. Finally, bulk measurements with $B$//[110] would be indispensable to guide the neutron diffraction experiments towards certain regions of the $B$-$T$ phase diagram.

**Acknowledgements**

The experiments of E. G. M at BENSC in Berlin, were supported by the European Commission under the Access to Research Infrastructure Action of the Improving Human Potential Programme (contract: HPRI-CT-1999-00020). E. G. M. also thanks Dr. P. Smeibidl for his assistance with the dilution refrigerator. She is also grateful to Dr. N. Stüßer and Dr. J. Hernandez-Velasco for many clarifying discussions and for their contribution to the preliminary experiment on the E6 at BENSC. Work at Los Alamos National Laboratory was performed under the auspices of the US Department of Energy.

FIGURE CAPTIONS

Fig. 1. Magnetic field dependence of the integrated intensity of the magnetic peak (1/2,1/2,1) at (a) 2.5 K, (b) 2.3 K, (c) 0.77 K and (d) 0.068 K.

Fig. 2. Magnetic field dependence of the integrated intensity of the reflection (1,1,1) at 2.5 K.

Fig. 3. Magnetic field dependence of the integrated intensity of the reflection (0,0,1) at (a) 2.5 K, and (b) 0.068 K.

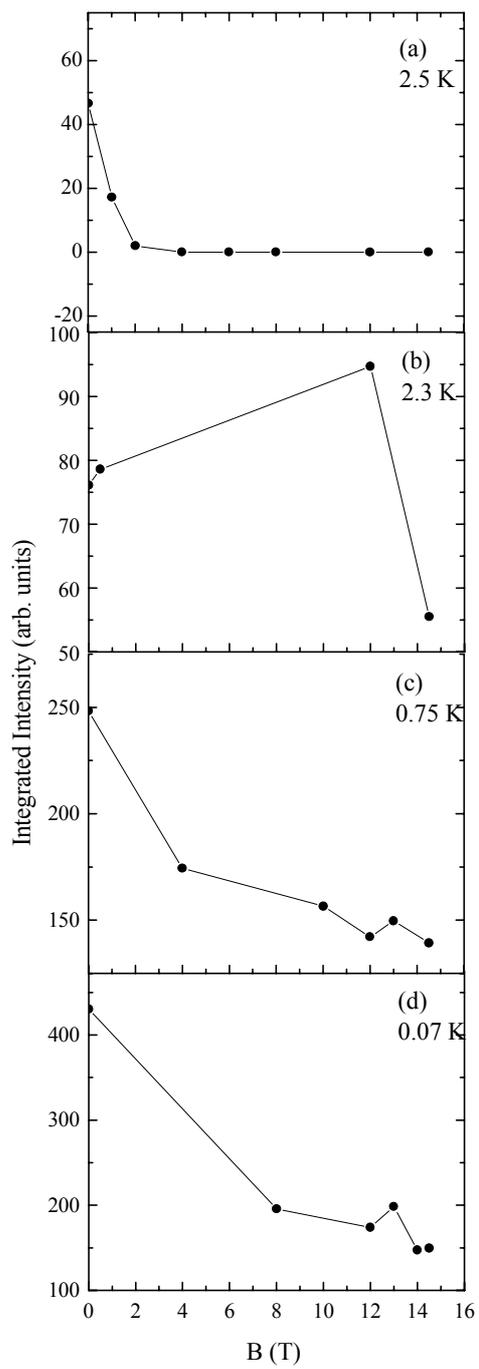

Fig. 1

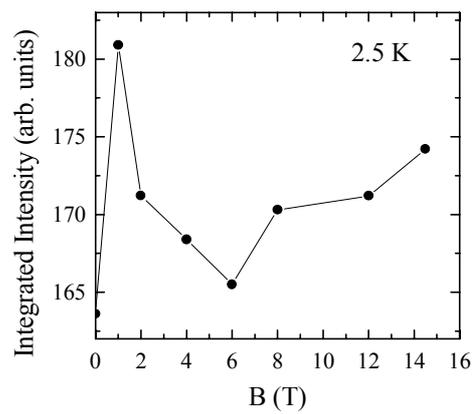

Fig. 2

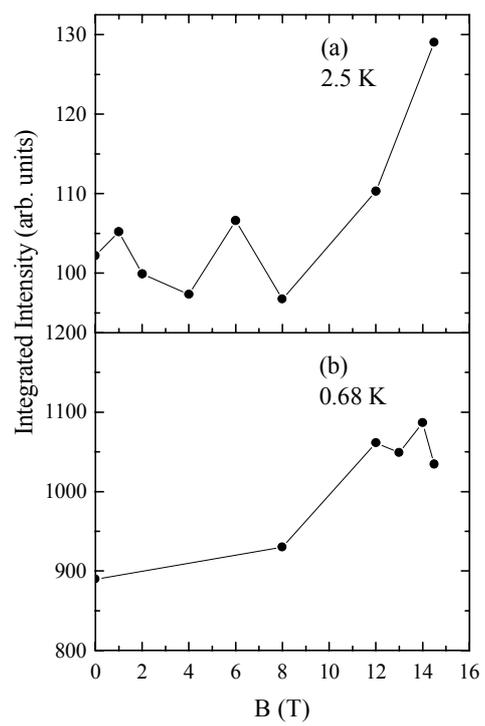

Fig. 3